\documentclass[twocolumn, journal]{IEEEtran}

\IEEEoverridecommandlockouts

\usepackage{color}
\usepackage{graphicx}
\usepackage{amsmath}
\usepackage{amssymb}
\usepackage{algorithm}
\usepackage{algorithmic}
\usepackage{amsmath}
\usepackage{multirow}
\usepackage{booktabs}
\usepackage{array}
\usepackage{amsthm}
\usepackage{mathrsfs}
\usepackage{stfloats}
\usepackage{caption}
\usepackage{subfigure}
\usepackage{bm}
\usepackage{setspace}
\usepackage{lineno}

\newcommand{\be}{\begin{equation}}
\newcommand{\ee}{\end{equation}}
\newcommand{\bea}{\begin{eqnarray}}
\newcommand{\eea}{\end{eqnarray}}
\newcommand{\ba}{\begin{array}}
\newcommand{\ea}{\end{array}}
\newcommand{\nid}{\noindent}

\allowdisplaybreaks[4]

\title{RIS-Aided Integrated Sensing and Communication: Joint Beamforming and Reflection Design
\thanks{H. Luo, R. Liu, and M. Li are with the School of Information and Communication Engineering, Dalian University of Technology, Dalian 116024, China (e-mail: luohonghao@mail.dlut.edu.cn; liurang@mail.dlut.edu.cn; mli@dlut.edu.cn).}
\thanks{Q. Liu is with the School of Computer Science and Technology, Dalian University of Technology, Dalian 116024, China (e-mail: qianliu@dlut.edu.cn).}
}

\author{Honghao Luo,
        Rang Liu,~\IEEEmembership{Graduate Student Member,~IEEE,}
        Ming Li,~\IEEEmembership{Senior Member,~IEEE,}\\
        and Qian Liu,~\IEEEmembership{Member,~IEEE}
        }

\begin{document}
\maketitle
\begin{abstract}
Integrated sensing and communication (ISAC) has been envisioned as a promising technique to alleviate the spectrum congestion problem.
Inspired by the applications of reconfigurable intelligent surface (RIS) in dynamically manipulating wireless propagation environment, in this paper, we investigate to deploy a RIS in an ISAC system to pursue performance improvement.
Particularly, we consider a RIS-assisted ISAC system where a multi-antenna base station (BS) performs multi-target detection and multi-user communication with the assistance of a RIS.
Our goal is maximizing the weighted summation of target detection signal-to-noise ratios (SNRs) by jointly optimizing the transmit beamforming and the RIS reflection coefficients, while satisfying the communication quality-of-service (QoS) requirement, the total transmit power budget, and the restriction of RIS phase-shift.
An efficient alternating optimization algorithm combining the majorization-minimization (MM), penalty-based, and manifold optimization methods is developed to solve the resulting complicated non-convex optimization problem.
Simulation results illustrate the advantages of deploying RIS in ISAC systems and the effectiveness of our proposed algorithm.
\end{abstract}


\section{Introduction}
\thispagestyle{empty}
With the explosive growth in demands of wireless communication, spectrum resources are becoming increasingly scare.
Researchers have recently engaged in investigating the spectrum sharing between radar systems and communication systems, which is known as the integrated sensing and communication (ISAC) technique.
ISAC enables a fully-shared platform that leverages shared resources to simultaneously implement communication and sensing functionalities, which greatly promotes the spectrum and hardware efficiencies.
Therefore, ISAC has been envisioned as one of the key enabling technologies in the future network \cite{R. Liu}, \cite{F. Liu}.

One of the major research fields in ISAC systems is designing the dual-functional transmit waveform.
Since the multi-input multi-output (MIMO) technique can exploit more spatial degrees-of-freedom (DoFs), it has been widely employed in ISAC systems to achieve beamforming gains \cite{Liu JSAC 2022}.
However, severe channel degradation is inevitable in practical environment, resulting in the unsatisfactory performance.
One promising technology to tackle this problem is the reconfigurable intelligent surface (RIS) technique.

RIS is generally composed of a planar metasurface with passive, cost-effective, and hardware-efficient reflecting elements \cite{Liu JSTSP 2021}.
Since additional non-line-of-sight (NLoS) links are established by deploying the RIS, more DoFs can be exploited to promote the system performance.
In light of these advantages, the application of RIS in ISAC systems has attracted extensive research \cite{Zhang 2022}-\cite{Y. Li}.
In \cite{Zhang 2022}, the authors jointly optimized the transmit beamforming and the RIS elements to maximize the communication data rate and the mutual information (MI) for radar sensing.
The authors in \cite{Song WCNC 2022} deployed a RIS in an ISAC system with multiple targets and single user.
A similar scenario was also considered in \cite{Y. Li} where a dual-function base station (BS) simultaneously detects single target and serves multiple users with the assistance of a RIS.
It is noticed that both \cite{Song WCNC 2022} and \cite{Y. Li} focused on the special scenario where the direct link between the BS and the target is blocked.
Moreover, the general scenario with multiple potential targets and multiple users is not considered in these works.

Motivated by the above findings, in this paper we consider a more typical scenario where a dual-functional BS successively detects multiple point-like targets and simultaneously serves multiple users via the direct links as well as with the assistance of a RIS.
In particular, we aim to jointly optimize the transmit beamforming and RIS elements to promote the target detection performance, in terms of the weighted summation of radar signal-to-noise ratios (SNRs), under the constraints of communication quality-of-service (QoS) requirement, the total transmit power budget, and the restriction of RIS elements.
In order to tackle the resulting complicated non-convex optimization problem, we develop an efficient alternating optimization algorithm combining the majorization-minimization (MM), penalty-based, and manifold optimization methods.
Finally, numerical results reveal the advantages of deploying RIS in ISAC systems and the effectiveness of our proposed algorithm.

\section{System Model and Problem Formulation}
\begin{figure}[t]
\centering
\includegraphics[width=2.8 in]{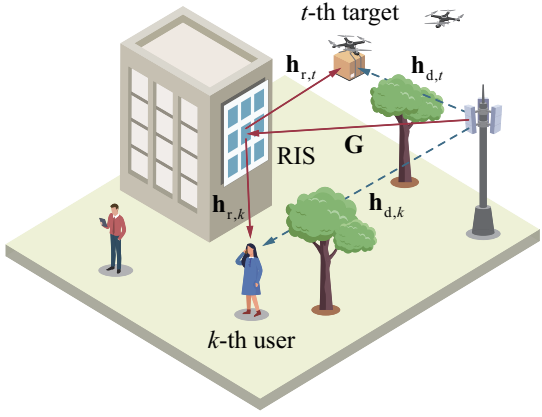}
\caption{A RIS-assisted ISAC system.}
\vspace{-0.2 in}
\label{fig:system_model}
\end{figure}

We consider a narrowband RIS-assisted ISAC system as depicted in Fig. \ref{fig:system_model}, where a multi-antenna BS equipped with $M$ transmit/receive antennas detects $T$ point-like targets and simultaneously serves $K$ single-antenna users with the assistance of an $N$-element RIS.
In order to avoid the strong interference from different targets and reduce the design complexity, $T$ potential targets are successively detected in a time-division fashion during the channel coherent time of communication users.
In specific, the transmission period is divided into $T$ time-slots, each of which is utilized to detect one target and simultaneously serves $K$ users.

The transmit signal of the BS is expressed as \cite{{Liu TSP 2020}}
\begin{equation}
\mathbf{x}=\mathbf{W}_\mathrm{c}\mathbf{s}_\mathrm{c}
+\mathbf{W}_\mathrm{r}\mathbf{s}_\mathrm{r}=\mathbf{W}\mathbf{s},
\end{equation}
where $\mathbf{W}_\mathrm{c}\in\mathbb{C}^{M \times K}$ and $\mathbf{W}_\mathrm{r}\in\mathbb{C}^{M \times M}$ denote the communication beamforming matrix and the radar beamforming matrix, respectively.
Vector $\mathbf{s}_\mathrm{c} \in \mathbb{C}^{K}$ represents the communication symbol vector with $\mathbb{E}\{\mathbf{s}_\mathrm{c}\mathbf{s}_\mathrm{c}^H\}=\mathbf{I}_K$, and vector $\mathbf{s}_\mathrm{r} \in \mathbb{C}^{M}$ is the radar probing signal with $\mathbb{E}\{\mathbf{s}_\mathrm{r}\mathbf{s}_\mathrm{r}^H\}= \mathbf{I}_M$.
They are assumed to be statistically independent of each other.
For brevity, we define the overall beamforming matrix as $\mathbf{W}\triangleq\left[\mathbf{W}_\mathrm{c}~ \mathbf{W}_\mathrm{r}\right] \in \mathbb{C}^{M \times (K+M)}$ and the transmit symbol vector as $\mathbf{s}\triangleq\left[\mathbf{s}_\mathrm{c}^T~ \mathbf{s}_\mathrm{r}^T\right]^T\in \mathbb{C}^{K+M}$.

From the radar detection perspective, since the target detection probability is proportional to the radar SNR, we aim to maximize the weighted radar sum-SNR of all targets to pursue better target detection performance.
As shown in Fig. \ref{fig:system_model}, the transmit probing signal will reach the $t$-th target via the direct link (the blue dashed line) and the reflected link (the red solid line), and then reflect back to the BS via all propagation links.
The propagation delay between these paths is negligible and not taken into consideration.
Thus, during the $t$-th time-slot, the collected echo signal from the $t$-th target is given by
\begin{align}
\mathbf{y}_{\mathrm{r},t}=(\mathbf{h}_{\mathrm{d},t}+\mathbf{G}^T\bm{\Phi}\mathbf{h}_{\mathrm{r},t})
(\mathbf{h}_{\mathrm{d},t}^T+\mathbf{h}_{\mathrm{r},t}^T\bm{\Phi}\mathbf{G})\mathbf{W}\mathbf{s}+\mathbf{n}_\mathrm{r},
\end{align}
where $\mathbf{h}_{\mathrm{d},t}\in\mathbb{C}^{M}, \mathbf{h}_{\mathrm{r},t}\in\mathbb{C}^{N}$ and $\mathbf{G}\in\mathbb{C}^{N\times M}$ denote the baseband channels between the BS and the $t$-th target, between the RIS and the $t$-th target, and between the BS and the RIS, respectively.
The target channel is the line-of-sight (LoS) link with given angle of departure (AoD).
The reflection matrix of the RIS is defined as $\bm{\Phi}\triangleq\mathrm{diag}(\bm{\phi})$ where $\bm{\phi}\triangleq[\phi_1,\cdots,\phi_N]^T$ is restricted with $|\phi_n|=1,\,\forall n$.
The vector $\mathbf{n}_\mathrm{r}\sim{\mathcal{C}\mathcal{N}(\mathbf{0},\sigma_\mathrm{r}^2\mathbf{I}_M})$ denotes the additive white Gaussian noise (AWGN).
Then, the radar SNR of the $t$-th target is thus given as
\begin{align}
\mathrm{SNR}_t=\frac{\mathbb{E}\big\{\left|\mathbf{H}_t\mathbf{W}\mathbf{s}\right|^2\big\}}{\sigma_\mathrm{r}^2}
=\frac{\mathrm{Tr}(\mathbf{W}^H\mathbf{H}_t^H\mathbf{H}_t\mathbf{W})}{\sigma_\mathrm{r}^2},
\end{align}
where for brevity we define the equivalent channel matrix as
\begin{align}
\label{eq:Ht}
\mathbf{H}_t\triangleq(\mathbf{h}_{\mathrm{d},t}+\mathbf{G}^T\bm{\Phi}\mathbf{h}_{\mathrm{r},t})
(\mathbf{h}_{\mathrm{d},t}^T+\mathbf{h}_{\mathrm{r},t}^T\bm{\Phi}\mathbf{G}).
\end{align}

From the communication perspective, a widely adopted criterion is guaranteeing the communication SINR requirement.
The received signal at the $k$-th user can be expressed as
\begin{align}
y_k = ( \mathbf{h}_{\mathrm{d},k}^T + \mathbf{h}_{\mathrm{r},k}^T\bm{\Phi}\mathbf{G} )\mathbf{x}+n_k,
\end{align}
where $\mathbf{h}_{\mathrm{d},k}\!\!\,\in\!\mathbb{C}^{M}$ and $\mathbf{h}_{\mathrm{r},k}\!\!\,\in\!\mathbb{C}^{N}$ denote the baseband channels between the BS and the $k$-th user, and between the RIS and the $k$-th user, respectively.
It is assumed that all the channel state information (CSI) is perfectly known at the BS given efficient channel estimation approaches \cite{Yang 2022}.
The scalar $n_k\!\sim{\mathcal{C}\mathcal{N}(0,\sigma_k^2)}$ denotes the AWGN with variance $\sigma_k^2$ at the $k$-th user.
The communication SINR of the $k$-th user is thus given as
\begin{align}
\mathbf{\gamma}_k = \frac{| ( \mathbf{h}_{\mathrm{d},k}^T+\mathbf{h}_{\mathrm{r},k}^T\bm{\Phi}\mathbf{G} )\mathbf{w}_k |^2}{\sum_{j \neq k}^{K+M}
| ( \mathbf{h}_{\mathrm{d},k}^T+\mathbf{h}_{\mathrm{r},k}^T\bm{\Phi}\mathbf{G})\mathbf{w}_j |^2+\sigma_k^2},
\end{align}
where $\mathbf{w}_j$ represents the $j$-th column of the beamforming matrix $\mathbf{W},j=1,\cdots,K+M$.

In this paper, we jointly design the beamforming matrix $\mathbf{W}$ and the reflection coefficients $\bm{\phi}$ to maximize the weighted radar sum-SNR, while satisfying the communication QoS requirement, the total transmit power budget, and the unit-modulus phase-shift of the RIS.
The optimization problem can be formulated as
\begin{subequations}
\label{eq:initial_problem}
\begin{align}
\max_{\mathbf{W},\bm{\phi}}~~&\sum_{t=1}^{T}\omega_t\mathrm{SNR}_t\\
\text{s.t.}~~&\gamma_k \geq \Gamma_k,~\forall k,\\
&\| \mathbf{W} \|_F^2 \leq P,\\
&\left| \phi_n \right|=1,~\forall n,
\end{align}
\end{subequations}
where $\omega_t$ represents the weighted coefficient of the $t$-th target, $\Gamma_k$ denotes the pre-defined communication SINR requirement of the $k$-th user, and $P$ is the total transmit power.
We observe that problem (\ref{eq:initial_problem}) is a complicated non-convex problem, which is difficult to optimize due to the non-convex objective function (\ref{eq:initial_problem}a), the unit-modulus constraint (\ref{eq:initial_problem}d), and the mutually coupled variables $\mathbf{W}$ and $\bm{\phi}$.
Thus, an alternating optimization strategy is exploited to solve problem (\ref{eq:initial_problem}) in the next section.

\section{Joint Beamforming and RIS Reflection Design For RIS-Assisted ISAC Systems}
In this section, we convert problem (\ref{eq:initial_problem}) into two more tractable sub-problems and develop efficient algorithms to iteratively solve them.

\subsection{Optimize the beamforming matrix $\mathbf{W}$}
With given $\bm{\phi}$ and some algebra transformations, the optimization problem with respect to $\mathbf{W}$ can be expressed as
\begin{subequations}
\label{eq:opt_W1}
\begin{align}
\max_{\mathbf{W}}~~&\mathrm{Tr}(\mathbf{W}^H\mathbf{C}_1\mathbf{W})\\
\text{s.t.}~~&(1\!+\!\Gamma_k^{-1})\mathbf{h}_k^T\mathbf{w}_k\mathbf{w}_k^H\mathbf{h}_k^*\!\geq\! \mathbf{h}_k^T\mathbf{W}\mathbf{W}^H\mathbf{h}_k^*\!+\!\sigma_k^2,\forall k,\\
&\| \mathbf{W} \|_F^2 \leq P,
\end{align}
\end{subequations}
where for simplicity we define
\begin{align}
\label{eq:hk}
\mathbf{h}_k^T&\triangleq\mathbf{h}_{\mathrm{d},k}^T+\mathbf{h}_{\mathrm{r},k}^T\bm{\Phi}\mathbf{G},\\
\mathbf{C}_1&\triangleq\sum_{t=1}^{T}\frac{\omega_t\mathbf{H}_t^H\mathbf{H}_t}{\sigma_r^2}.
\end{align}
It is obvious that problem (\ref{eq:opt_W1}) is a non-convex quadratically constrained quadratic programming (QCQP) problem, which can be converted into a semi-definite programming (SDP) problem by utilizing the semi-definite relaxation (SDR) strategy.
However, a rank-one optimal solution usually cannot be guaranteed.
In this case, some approximation techniques such as Gaussian randomization should be employed to construct an approximately optimal solution, which dramatically increases the computational complexity and cannot guarantee the convergence of overall algorithm.

In order to avoid the prohibitively high complexity of the SDR strategy, we propose to construct a series of easy-to-optimize surrogate functions for the non-convex objective function (\ref{eq:opt_W1}a) based on the MM method.
Specifically, with the obtained solution $\mathbf{W}^i$ in the $i$-th iteration, a more tractable surrogate function, which approximates (\ref{eq:opt_W1}a) at the current local point $\mathbf{W}^i$ and serves
as a lower bound, is constructed as
\begin{align}
\begin{split}
\label{eq:opt_W_MM}
\mathrm{Tr}(\mathbf{W}^H\mathbf{C}_1\mathbf{W})
&=\widetilde{\mathbf{w}}^H(\mathbf{C}_1^T\otimes\mathbf{I})\widetilde{\mathbf{w}}\\
&\geq 2\Re\{(\widetilde{\mathbf{w}}^i)^H(\mathbf{C}_1^T\otimes\mathbf{I})\widetilde{\mathbf{w}}\}+c_1,
\end{split}
\end{align}
where for simplicity we define
\begin{align}
\widetilde{\mathbf{w}}&\triangleq\mathrm{vec}(\mathbf{W}^H),\\
c_1&\triangleq-(\widetilde{\mathbf{w}}^i)^H(\mathbf{C}_1^T\otimes\mathbf{I})\widetilde{\mathbf{w}}^i.
\end{align}

Substituting (\ref{eq:opt_W_MM}) into (\ref{eq:opt_W1}a) and ignoring the constant term, the beamforming design problem (\ref{eq:opt_W1}) can be reformulated as
\begin{subequations}
\label{eq:opt_W2}
\begin{align}
\max_{\mathbf{W}}~&\Re\big\{{(\widetilde{\mathbf{w}}^i)^H(\mathbf{C}_1^T\otimes\mathbf{I})\widetilde{\mathbf{w}}}\big\}\\
\text{s.t.}\,~&(1\!+\!\Gamma_k^{-1})\mathbf{h}_k^T\mathbf{w}_k\mathbf{w}_k^H\mathbf{h}_k^*\!\geq\! \mathbf{h}_k^T\mathbf{W}\mathbf{W}^H\mathbf{h}_k^*\!+\!\sigma_k^2,\forall k,\\
&\| \mathbf{W} \|_F^2 \leq P.
\end{align}
\end{subequations}
Although the constraint (\ref{eq:opt_W2}b) is still non-convex, it can be converted into a form of second-order cone, which can be effectively solved by the second-order cone programming (SOCP) method \cite{Luo JSAC 2006}.

\subsection{Optimize the reflection coefficients $\bm{\phi}$}
With given $\mathbf{W}$, the optimization problem with respect to $\bm{\phi}$ can be formulated as
\begin{subequations}
\label{eq:opt_phi1}
\begin{align}
\max_{\bm{\phi}}~~&\sum_{t=1}^{T}\mathrm{vec}^H(\mathbf{H}_t)\mathbf{C}_{2,t}\mathrm{vec}(\mathbf{H}_t)\\
\mathrm{s.t.}~~\;&\frac{\left|\mathbf{h}_k^T\mathbf{w}_k\right|^2}{\sum_{j\neq k}^{K+M}\left|\mathbf{h}_k^T\mathbf{w}_j\right|^2+\sigma_k^2} \geq \Gamma_k,~\forall k,\\
&\left|\phi_n\right|=1,
\end{align}
\end{subequations}
where for brevity we define $\mathbf{C}_{2,t}\triangleq\omega_t\frac{(\mathbf{W}^*\mathbf{W}^T)\otimes\mathbf{I}}{\sigma_\mathrm{r}^2}$.
It is noted that the matrix $\mathbf{H}_t$ is quadratic with respect to $\bm{\phi}$ as denoted in (\ref{eq:Ht}), and thus the objective function (\ref{eq:opt_phi1}a) is a quartic function, which causes significant difficulties in solving problem (\ref{eq:opt_phi1}).
Moreover, the fractional terms with quadratic functions in constraint (\ref{eq:opt_phi1}b) and the unit-modulus constraint (\ref{eq:opt_phi1}c) also hinder the solution of problem (\ref{eq:opt_phi1}).

In order to tackle these difficulties, we construct a penalized problem and solve it with the MM and manifold methods.
Firstly, auxiliary variables $a_{k,j},\forall k,\forall j$ are introduced to transform the constraint (\ref{eq:opt_phi1}b) into
\begin{subequations}
\label{eq:penalty_term}
\begin{align}
&\frac{\left|a_{k,k}\right|^2}{\sum_{j\neq k}^{K+M}\left|a_{k,j}\right|^2+\sigma_k^2} \geq \Gamma_k,~\forall k,\\
&a_{k,j}=\mathbf{h}_k^T\mathbf{w}_j,~\forall k,~\forall j.
\end{align}
\end{subequations}
Then, constraint (\ref{eq:penalty_term}b) is relaxed and added to the objective function as a penalty term.
Thus, the original problem (\ref{eq:opt_phi1}) can be converted into the following penalized problem
\begin{subequations}
\label{eq:opt_phi2}
\begin{align}
\min_{\bm{\phi},\;{a_{k,j}},\forall k,\forall j}&-\!\!\sum_{t=1}^{T}\mathbf{v}_t^H\mathbf{C}_{2,t}\mathbf{v}_t\!+\!
\rho\!\!\sum_{k=1}^{K}\!\!\sum_{j=1}^{K+M}\!\!\left|a_{k,j}\!-\!\mathbf{h}_k^T\mathbf{w}_j\right|^2\\
\mathrm{s.t.}~~~\,\,&\,\frac{\left|a_{k,k}\right|^2}{\sum_{j\neq k}^{K+M}\left|a_{k,j}\right|^2+\sigma_k^2} \geq \Gamma_k,~\forall k,\\
&\left|\phi_n\right|=1,
\end{align}
\end{subequations}
where $\rho>0$ represents the penalty coefficient and for brevity we define
\begin{align}
\mathbf{B}_t\triangleq&~\mathbf{G}^T\mathrm{diag}\{\mathbf{h}_{\mathrm{r},t}\},\\
\begin{split}
\label{eq:v}
\mathbf{v}_t\triangleq&~\mathrm{vec}(\mathbf{H}_t)\\
=&~(\mathbf{B}_t\otimes\mathbf{B}_t)\mathrm{vec}\{\bm{\phi}\bm{\phi}^T\}
+\mathrm{vec}(\mathbf{h}_{\mathrm{d},t}\mathbf{h}_{\mathrm{d},t}^T)\\
&+(\mathbf{h}_{\mathrm{d},t}\otimes\mathbf{B}_t+\mathbf{B}_t\otimes\mathbf{h}_{\mathrm{d},t})\bm{\phi}.
\end{split}
\end{align}

With given $\rho$, since the variables $\bm{\phi}$ and $a_{k,j},\forall k,\forall j$ are coupled in problem (\ref{eq:opt_phi2}), we propose to convert it into two sub-problems and iteratively optimize each variable as follows.

With given $a_{k,j},\forall k,\forall j$, the optimization problem (\ref{eq:opt_phi2}) with respect to $\bm{\phi}$ can be formulated as
\begin{subequations}
\label{eq:opt_phi3}
\begin{align}
\min_{\bm{\phi}}~~&-\!\sum_{t=1}^{T}\mathbf{v}_t^H\mathbf{C}_{2,t}\mathbf{v}_t+ \rho\sum_{k=1}^{K}\!\!\sum_{j=1}^{K+M}\left|a_{k,j}-\mathbf{h}_k^T\mathbf{w}_j\right|^2\\
\mathrm{s.t.}~~\;&\left|\phi_n\right|=1.
\end{align}
\end{subequations}
It can be easily observed that problem (\ref{eq:opt_phi3}) is still difficult to optimize due to the non-convex objective function (\ref{eq:opt_phi3}a) and unit-modulus constraint (\ref{eq:opt_phi3}b).
Although the objective function (\ref{eq:opt_phi3}a) is smooth and differentiable, the first term $-\sum_{t=1}^{T}\mathbf{v}_t^H\mathbf{C}_{2,t}\mathbf{v}_t$ is a complicated quartic polynomial expression with respect to $\bm{\phi}$, whose first-order derivative is difficult to obtain.
Thus, we exploit the MM method to make it easier to tackle.
Specifically, with the obtained solution $\bm{\phi}^i$ and $\mathbf{v}_t^i$ in the $i$-th iteration, an upper bound of the first term in (\ref{eq:opt_phi3}a) can be constructed as
\begin{align}
\label{eq:opt_phi_MM}
-\!\!\sum_{t=1}^{T}\!\mathbf{v}_t^H\mathbf{C}_{2,t}\mathbf{v}_t\!\leq\!
-\!\!\sum_{t=1}^{T}\!\!\Big(\!2\Re\big\{(\mathbf{v}_t^i)^H\mathbf{C}_{2,t}\mathbf{v}_t\!\big\}
\!-\!\!\,(\mathbf{v}_t^i)^H\mathbf{C}_{2,t}\mathbf{v}_t^i\Big).
\end{align}

Then, utilizing the definition of $\mathbf{v}_t$ in (\ref{eq:v}), we have
\begin{small}
\begin{align}
\begin{split}
\label{eq:v2phi}
\sum_{t=1}^{T}2\Re\big\{(\mathbf{v}_t^i)^{H}\mathbf{C}_{2,t}\mathbf{v}_t\big\}
\!&=\!\,\Re\big\{(\mathbf{f}_1^i)^H\!\mathrm{vec}\{\bm{\phi}\bm{\phi}^T\!\}\!
+\!(\mathbf{f}_2^i)^H\bm{\phi}\big\}\!+\!c_2\\[-0.05 in]
&=\!\,\Re\big\{\bm{\phi}^H\mathbf{F}_1^i\bm{\phi}^*\!+\!(\mathbf{f}_2^i)^H\bm{\phi}\big\}\!+\!c_2,
\end{split}
\end{align}
\end{small}

\nid where the matrix $\mathbf{F}_1^i\in\mathbb{C}^{N\times N}$ is a reshaped version of the vector $\mathbf{f}_1^i$, i.e., $\mathbf{f}_1^i=\mathrm{vec}\{\mathbf{F}_1^i\}$, and for simplicity we define
\begin{align}
\mathbf{f}_1^i\triangleq&~\sum_{t=1}^{T}2(\mathbf{B}_t^H\otimes\mathbf{B}_t^H)\mathbf{C}_{2,t}\mathbf{v}_t^i,\\
\mathbf{f}_2^i\triangleq&~\sum_{t=1}^{T}2(\mathbf{B}_t^H\!\otimes\mathbf{h}_{\mathrm{d},t}^H
+\mathbf{h}_{\mathrm{d},t}^H\otimes\mathbf{B}_t^H)
\mathbf{C}_{2,t}\mathbf{v}_t^i,\\
c_2\triangleq&~\sum_{t=1}^{T}2\Re\big\{(\mathbf{v}_t^i)^H\mathbf{C}_{2,t}
\mathrm{vec}(\mathbf{h}_{\mathrm{d},t}\mathbf{h}_{\mathrm{d},t}^T)\big\}.
\end{align}

Similarly, based on (\ref{eq:hk}) and some algebra transformations, the second term in (\ref{eq:opt_phi3}a) can be equivalently re-written as
\begin{align}
\label{eq:trans_obj}
f(\bm{\phi})\triangleq\rho(\bm{\phi}^T\mathbf{Q}\bm{\phi}^*+\mathbf{q}^T\bm{\phi}+c_3),
\end{align}
where for brevity we define
\begin{subequations}
\begin{align}
\mathbf{Q}&\triangleq\sum_{k=1}^{K}\!\!\sum_{j=1}^{K+M}\!\!
\mathrm{diag}\{\mathbf{h}_{\mathrm{r},k}\}\mathbf{G}\mathbf{w}_j
\mathbf{w}_j^H\mathbf{G}^H\mathrm{diag}\{\mathbf{h}_{\mathrm{r},k}^*\},\\
\mathbf{q}&\triangleq\sum_{k=1}^{K}\!\!\sum_{j=1}^{K+M}\!\!2\Re\big\{(\mathbf{w}_j^H\mathbf{h}_{\mathrm{d},k}^*\!-a_{k,j}^*)
\mathrm{diag}\{\mathbf{h}_{\mathrm{r},k}\}\mathbf{G}\mathbf{w}_j \big\},\\
c_3&\triangleq\sum_{k=1}^{K}\!\!\sum_{j=1}^{K+M}\!\!\big|a_{k,j}-\mathbf{h}_{\mathrm{d},k}^T\mathbf{w}_j\big|^2.
\end{align}
\end{subequations}

Substituting (\ref{eq:opt_phi_MM}), (\ref{eq:v2phi}) and (\ref{eq:trans_obj}) into (\ref{eq:opt_phi3}a), and ignoring constant terms, problem (\ref{eq:opt_phi3}) can be reformulated as
\begin{subequations}
\label{eq:opt_phi4}
\begin{align}
\min_{\bm{\phi}}~~&-\Re\big\{\bm{\phi}^H\mathbf{F}_1^i\bm{\phi}^*+(\mathbf{f}_2^i)^H\bm{\phi}\big\}+f(\bm{\phi})\\
\mathrm{s.t.}~~\;&\left|\phi_n\right|=1.
\end{align}
\end{subequations}
We observe that the objective function (\ref{eq:opt_phi4}a) is smooth with easy-to-obtain derivatives.
Additionally, the unit-modulus constraint (\ref{eq:opt_phi4}b) forms a complex circle Riemannian manifold, which allows problem (\ref{eq:opt_phi4}) to be solved by the typical Riemannian conjugate gradient (RCG) algorithm.
After deriving the Riemannian gradient from the corresponding Euclidean gradient, we utilize the idea of conjugate gradient algorithm to solve problem (\ref{eq:opt_phi4}) on the Riemannian space.
The details of RCG algorithm are omitted here due to space limitations and can be found in \cite{Rang TWC 2021}.

With given $\bm{\phi}$, the sub-problem of updating $a_{k,j},\forall k,\forall j$ can be formulated as
\begin{subequations}
\label{eq:opt_a}
\begin{align}
\min_{a_{k,j},\forall k,\forall j} ~~&\rho\sum_{k=1}^{K}\sum_{j=1}^{K+M}\left|a_{k,j}-\mathbf{h}_k^T\mathbf{w}_j\right|^2\\
\mathrm{s.t.}~~\;&\frac{\left|a_{k,k}\right|^2}{\sum_{j\neq k}^{K+M}\left|a_{k,j}\right|^2+\sigma_k^2} \geq \Gamma_k,~\forall k.
\end{align}
\end{subequations}
It is clear that problem (\ref{eq:opt_a}) can be effectively tackled with the SOCP method.

After updating $\mathbf{W}$, $\bm{\phi}$, and $a_{k,j},\forall k,\forall j$, the penalty coefficient $\rho$ is updated by
\begin{align}
\label{eq:opt_rho}
\rho=\frac{\rho}{c},\quad 0<c<1,
\end{align}
where $c$ is a constant parameter to gradually increase the value of $\rho$ for the sake of tightening the penalty term.
In order to find a good starting point for the proposed algorithm, the initial value of $\rho$ is set as a relatively small number.

\subsection{Summary}
Based on the aforementioned derivations, the proposed algorithm is summarized as follows.
With appropriate initial values \cite{Liu JSTSP 2021}, the beamforming matrix $\mathbf{W}$, the reflection coefficients $\bm{\phi}$, the auxiliary variables $a_{k,j},\forall k,\forall j$, and the penalty coefficient $\rho$ are iteratively updated by (\ref{eq:opt_W2}), (\ref{eq:opt_phi4}), (\ref{eq:opt_a}) and (\ref{eq:opt_rho}), respectively, until the achievable radar SNR (\ref{eq:initial_problem}a) converges as well as the stopping indicator $\zeta$ is below the threshold $\epsilon$.
Particularly, the stopping indicator $\zeta$ is defined as
\begin{align}
\zeta\triangleq\mathrm{max}\big\{|a_{k,j}-\mathbf{h}_k^T\mathbf{w}_j|,\forall k,\forall j\big\}.
\end{align}

\section{Simulation Results}
\begin{figure}[t]
\centering
  \includegraphics[width=2.8 in]{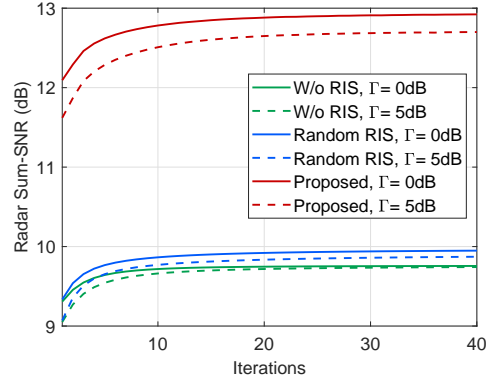}
  \caption{Convergence performance.}
  \label{fig:convergence}
\end{figure}

\begin{figure*}[htbp]
\centering
\begin{minipage}[t]{0.33\linewidth}
\setcaptionwidth{2.03 in}
\centering
\includegraphics[width=2.52 in,height=2 in]{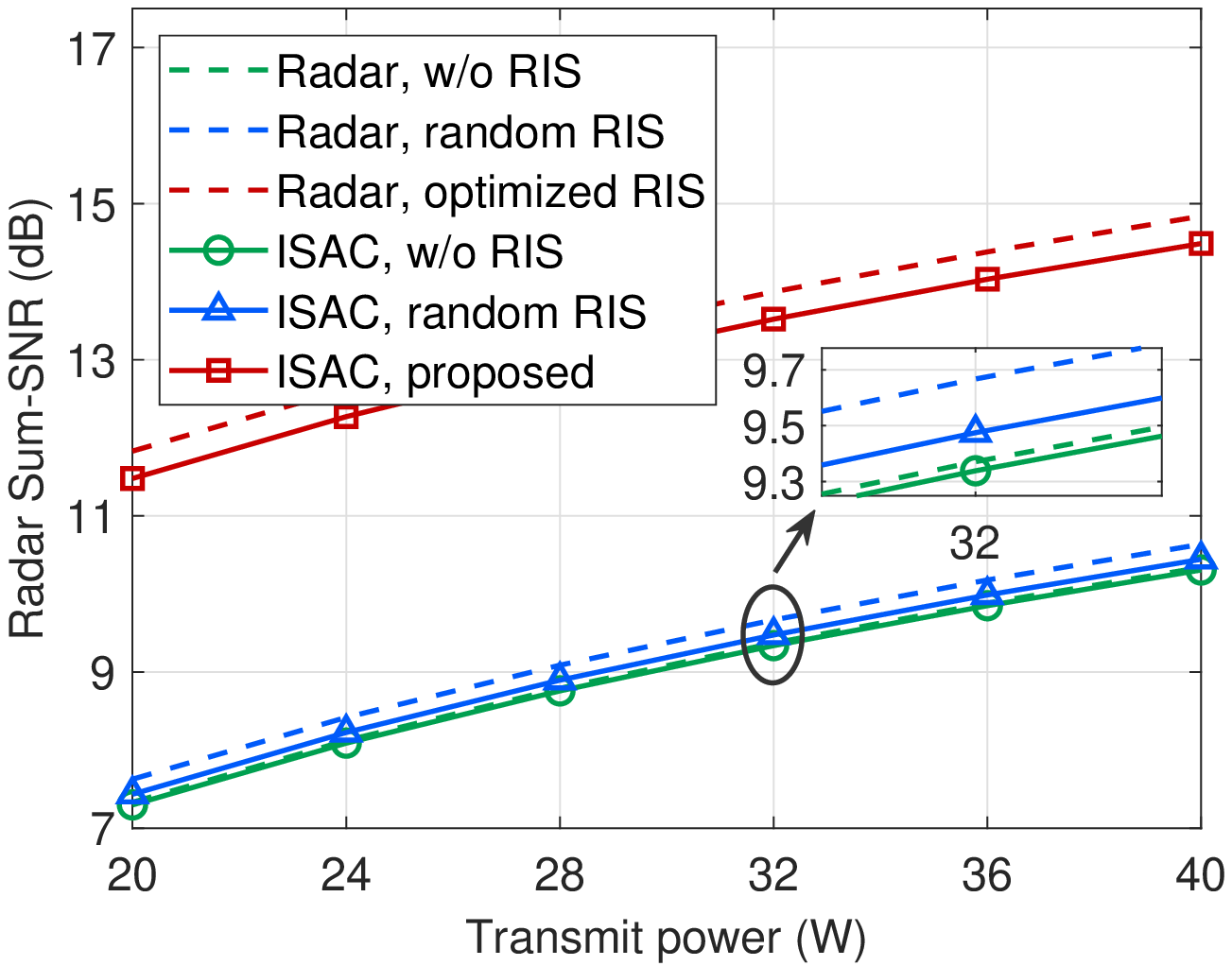}
\caption{Radar sum-SNR versus transmit power ($N=36,\,\Gamma=5\mathrm{dB}$).}
\label{fig:power}
\end{minipage}%
\begin{minipage}[t]{0.33\linewidth}
\setcaptionwidth{1.88 in}
\centering
\includegraphics[width=2.52 in,height=2 in]{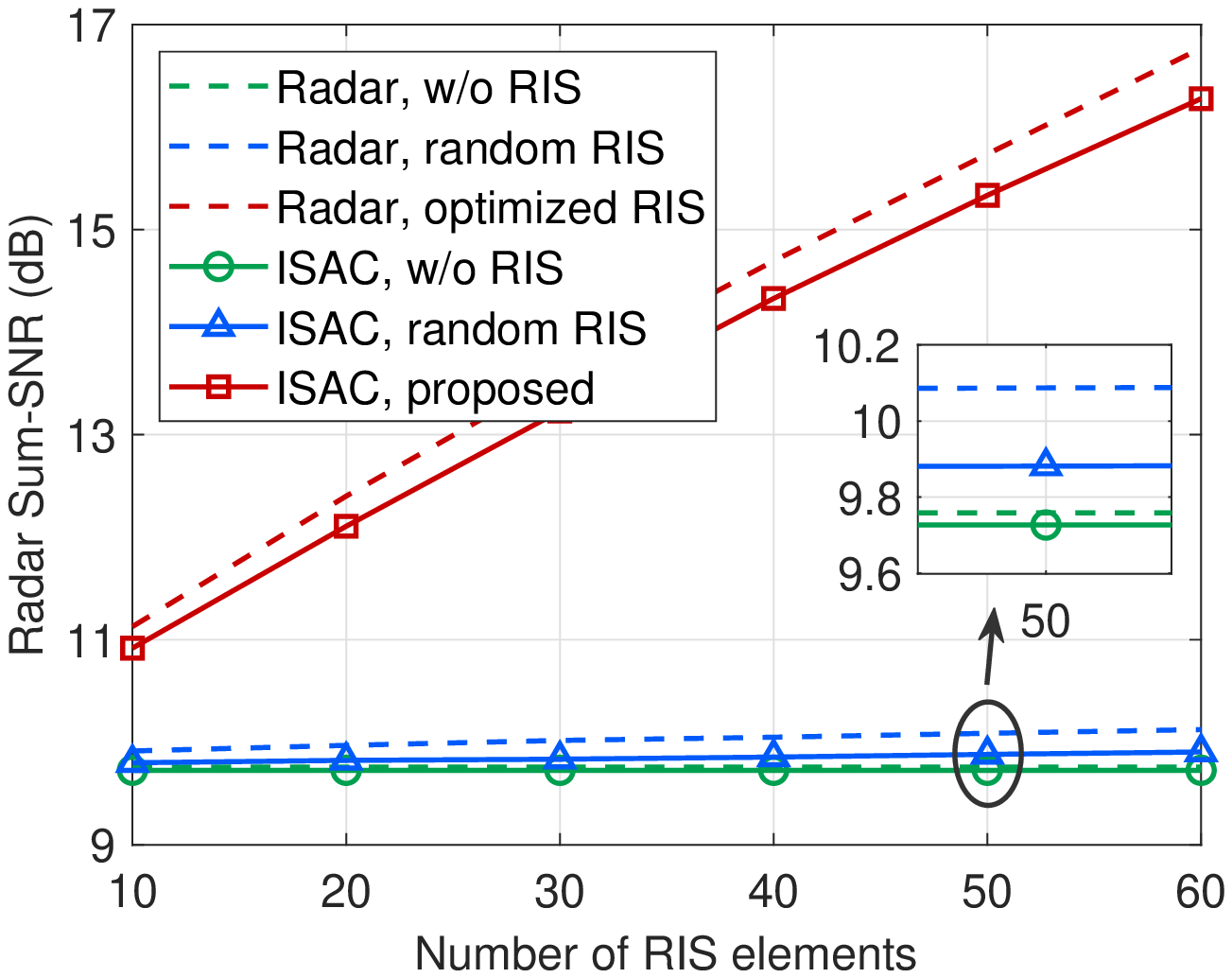}
\caption{Radar sum-SNR versus the number of RIS elements ($P=35\mathrm{W},\;\Gamma=5\mathrm{dB}$).}
\label{fig:RIS_elements}
\end{minipage}%
\begin{minipage}[t]{0.33\linewidth}
\setcaptionwidth{1.92 in}
\centering
\includegraphics[width=2.52 in,height=2 in]{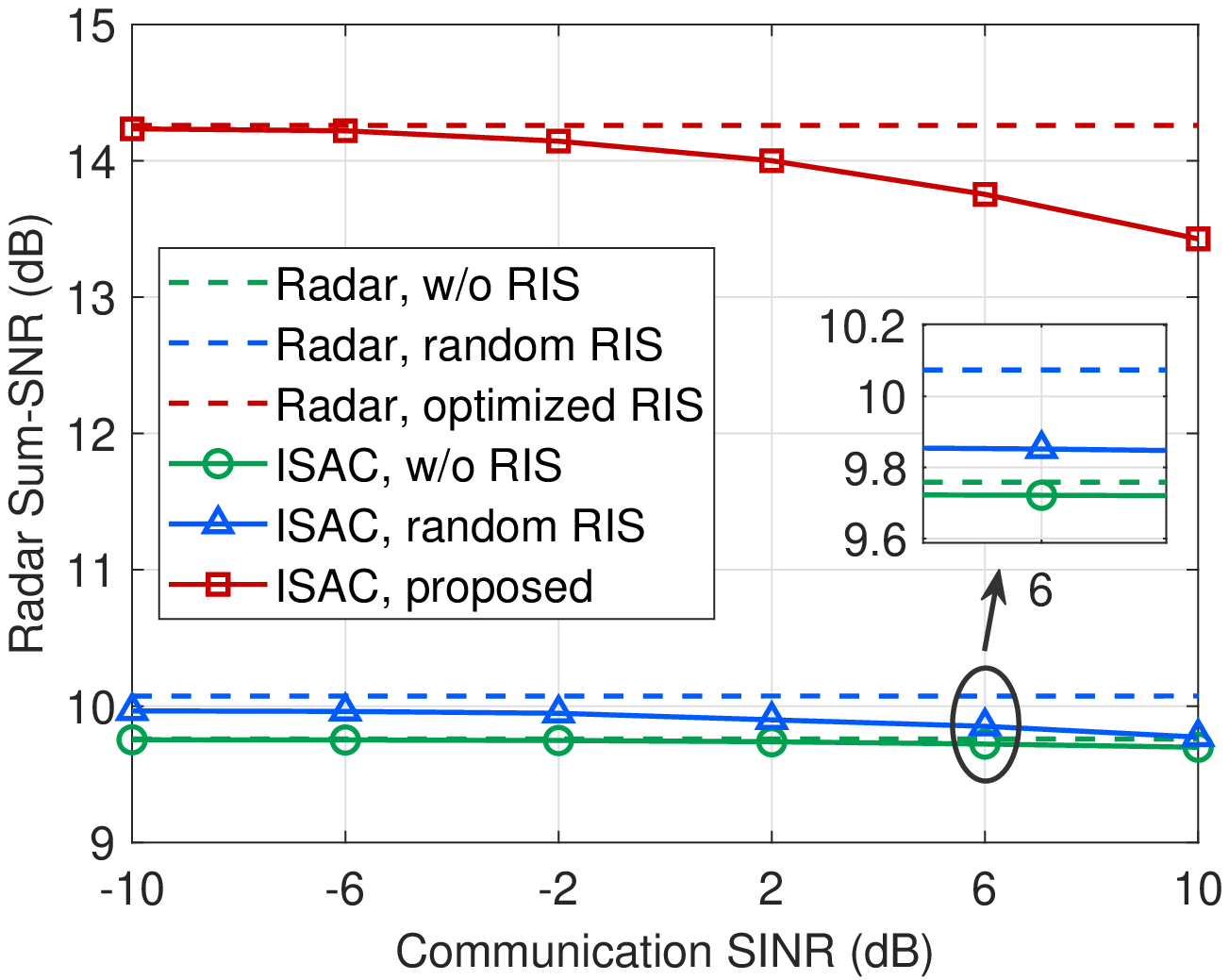}
\caption{Radar sum-SNR versus communication SINR requirement ($P=35\mathrm{W},\;N=36$).}
\label{fig:tradeoff}
\end{minipage}
\centering
\end{figure*}
In this section, we numerically evaluate the performance of our proposed algorithm in the considered RIS-assisted ISAC system.
It is assumed that the BS equipped with $M=16$ antennas in half-wavelength antenna spacing detects $T=3$ point-like targets and simultaneously serves $K=4$ single-antenna users with the assistance of an $N$-element RIS.
The potential targets are located at the azimuth angles of [$-30^\circ$, $0^\circ$, $30^\circ$], and the users are distributed around the RIS.
The distances of BS-target, BS-RIS, and RIS-user links are set as 30m, 35m, and 3m, respectively.
We adopt the typical distance-dependent path-loss model \cite{Wu TWC 2019} and set the path-loss exponents of BS-RIS, RIS-target, RIS-user, BS-target and BS-user links as $\alpha_\mathrm{BR}=\alpha_\mathrm{Rt}=\alpha_\mathrm{Ru}=2.3$, $\alpha_\mathrm{Bt}=2.7$ and $\alpha_\mathrm{Bu}=3.3$, respectively.
The Rician fading channel model is assumed with the Rician factors of RIS-user link being $\beta_\mathrm{Ru}=3\mathrm{dB}$ and other links being 0dB.
The noise power at the radar receiver and the $k$-th user are set as $\sigma_r^2=\sigma_k^2=-80\mathrm{dBm},\;\forall k$.
The SINR requirement for all users is set to the same value $\Gamma_k=\Gamma,\;\forall k$.
We set the weighted coefficients $\omega_t=1$, $\forall t$ for simplicity.

We first present the convergence performance of the proposed algorithm in Fig. \ref{fig:convergence}.
It can be easily observed that the achievable radar sum-SNR of all schemes can converge within about 30 iterations.

The radar sum-SNR versus the transmit power is shown in Fig. \ref{fig:power}.
In order to verify the effectiveness of the proposed algorithm (``ISAC, proposed''), the schemes without RIS (``ISAC, w/o RIS'') and with random phase-shift RIS (``ISAC, random RIS'') are included.
Besides, the radar-only (``Radar'') systems with or without RIS are also included as benchmarks to present the upper bound of target detection performance.
We observe that the schemes with RIS are better than those without RIS in both radar-only and ISAC systems, because RIS introduces additional NLoS links to achieve passive beamforming gain.
Moreover, the ``ISAC, proposed'' scheme can achieve about 4dB performance improvement than other schemes in the considered ISAC system, which verifies the effectiveness of the proposed algorithm.
Additionally, there exists a performance gap between the radar-only system and the considered ISAC system due to the trade-off between the communication and radar detection performance.

Then, we plot the radar sum-SNR versus the number of RIS elements in Fig. \ref{fig:RIS_elements}.
It can be found that the radar sum-SNR increases with the increasing number of RIS elements, because more reflection elements provide more spatial DoFs to enhance the performance.
In addition, compared with the ``ISAC, w/o RIS'' scheme, the proposed algorithm for the RIS with $N=60$ elements achieves about 6dB performance gain by smartly tailoring the propagation environments.

Finally, we show the radar sum-SNR and communication QoS requirement in Fig. \ref{fig:tradeoff}.
Not surprisingly, there exists a trade-off between the radar detection and communication performance.
As the communication QoS requirement increases, the radar sum-SNR achieved by the considered ISAC system decreases, because more resources are allocated to guarantee the communication QoS.

\section{Conclusion}
In this paper, we investigated the joint transmit beamforming and RIS reflection design for a RIS-assisted ISAC system, where the RIS is deployed to simultaneously assist multi-target detection and multi-user communication.
The radar sum-SNR was maximized under the constraints of communication SINR, the total transmit power budget, and the RIS reflection coefficients.
An efficient alternating optimization algorithm was developed to solve the resulting complicated non-convex problem.
Numerical results demonstrated the advantages of deploying RIS in ISAC systems as well as revealed the performance trade-off.
We will extend our work to more practical scenarios where there exist clutters, as well as the crucial fairness issue.

\end{document}